\documentclass[10pt,a4paper,reqno]{amsart} 
 \usepackage{graphicx}
 \usepackage{amsthm}
 \usepackage{amsmath}
\usepackage{multicol}
  \usepackage{amssymb}
 \usepackage{enumerate}
 \usepackage{hyperref}
  \usepackage{epigraph}
  \usepackage{marvosym}
    \usepackage{wasysym}
    \usepackage{latexsym}
 \usepackage[usenames,dvipsnames]{color}
\usepackage[usenames,dvipsnames]{pstricks}
 \usepackage{epsfig}
 \usepackage{pst-grad} 
 \usepackage{pst-plot}

\usepackage{fullpage}

\hypersetup{
    unicode=false,          
    pdftoolbar=true,        
    pdfmenubar=true,        
    pdffitwindow=false,     
    pdfstartview={FitH},    
    pdftitle={Quantitative Mode Stability for the Wave Equation on the Kerr--Newman Spacetime},    
    pdfauthor={D. Civin},     
    pdfcreator={Damon Civin},   
    colorlinks=true,       
    linkcolor=black,          
    citecolor=black,        
    filecolor=magenta,      
    urlcolor=blue           
}

\newtheorem{theorem}{Theorem}[section]
\newtheorem{lemma}[theorem]{Lemma}
\newtheorem{proposition}[theorem]{Proposition}

\newtheorem{definition}[theorem]{Definition}

\newenvironment{remark}[1][Remark]{\begin{trivlist}
\item[\hskip \labelsep {\bfseries #1}]}{\end{trivlist}}

\newcommand{\be}{\begin{equation}}
\newcommand{\ee}{\end{equation}}

\newcommand{\ind}{_{m\ell}^{(a\omega)}} 

\newcommand{\w}{\omega}

\newcommand{\cut}{{\mbox{\Rightscissors}}}

\newcommand{\ds}{\displaystyle}

\newcommand{\baa}{\begin{align}}
\newcommand{\eaa}{\end{align}}

\newcommand{\bea}{\begin{eqnarray*}}
\newcommand{\eea}{\end{eqnarray*}}

\newcommand{\nn}{\nonumber}

\newcommand{\mr}{\mathrm}

\newcommand{\bp}{\begin{proof}[{\textbf{Proof}}]}

\newcommand{\ep}{\end{proof}}

\newcommand{\abs}[1]{{\left|#1\right|}}
\newcommand{\set}[1]{\left\{ #1 \right\} }

\newcommand{\audr}{\abs{u'}^2 }
\newcommand{\au}{\abs{u}^2 }

\newcommand{\bt}{\begin{theorem}}
\newcommand{\et}{\end{theorem}}

\newcommand{\bl}{\begin{lemma}}
\newcommand{\el}{\end{lemma}}

\newcommand{\bd}{\begin{definition}}
\newcommand{\ed}{\end{definition}}

\newcommand{\bei}{\begin{itemize}}
\newcommand{\eei}{\end{itemize}}

\newcommand{\ben}{\begin{enumerate}[(i)]}
\newcommand{\een}{\end{enumerate}}

\newcommand{\dt}{\partial_t}
\newcommand{\dd}{\partial}

\newcommand{\Lang}{\mbox{$\Delta \mkern-11mu /$\,}}

\newcommand{\Hp}{\mathcal H^+}

\newcommand{\remarks}{\noindent\textbf{Remarks}\begin{enumerate}}

\begin{document}

\title{Quantitative Mode Stability for the Wave Equation on the Kerr--Newman Spacetime}

\author{Damon Civin}

\address{Cambridge Centre for Analysis, University of Cambridge, CB3 0WA, United Kingdom}

\email{ \href{mailto:dc509@cam.ac.uk}{dc509@cam.ac.uk}}

\begin{abstract}
Quantitative versions of mode stability type statements for the wave equation on Kerr--Newman black holes are proven for the full sub-extremal range ($a^2 + Q^2 < M^2$). The mode stability result on the real axis is then applied to prove integrated local energy decay for solutions restricted to a bounded frequency regime. This is an important element of the proof of unrestricted boundedness and decay statements, presented in a forthcoming companion paper.

\end{abstract}

\thanks{I would like to express my gratitude to my supervisor, Prof.~M.~Dafermos, for suggesting the problem and his advice and to Y.~Shlapentokh-Rothman for insightful discussions and comments on the manuscript. 
I am also grateful to M.~Joaris for her caring and patience and my family for their support and encouragement. My thanks also go to my fellow students for providing an atmosphere in which it is a pleasure to work. 
I am supported jointly by a Cambridge Commonwealth Trust scholarship and UK Engineering and Physical Sciences Research Council (EPSRC) grant EP/H023348/1. 
}

\maketitle

\tableofcontents

\pagestyle{plain}

\section{Introduction}

Spacetime outside of a stationary, charged, rotating black-hole is described by a member of the subextremal 
family of Kerr--Newman solutions of the Einstein electrovacuum equations. 

One of the most important open problems in general relativity is the nonlinear stability of the exterior Kerr(--Newman) metric (see \cite{K07}).  At present, the only global nonlinear stability result in the asymptotically flat setting is that for Minkowski space, proved by Christodoulou and Klainerman in \cite{CK93}. Following the philosophy of their monumental proof, the first step in the journey toward resolution of the nonlinear stability problem is the analysis of the linear stability problem using sufficiently robust techniques. The simplest such linear problem is that of the scalar wave equation
\be
\label{eq:wave0} 
\square_g \psi = 0,
\ee
which may be thought of as a ``poor-man's version'' of the linearised Einstein equations (taken around a subextremal Kerr--Newman metric $g$). 
Thus, the boundedness and decay in time of such $\psi$ on a Kerr--Newman background may be thought of as stability of this spacetime for linear scalar perturbations. 
This linear stability result is proved in this paper's companion, \cite{C14-1}, following the approach taken for the Kerr case by Dafermos, Rodnianski and Shlapentokh-Rothman in \cite{DR10} and \cite{DRS14}.
 
As in the Kerr case, one of the major difficulties in understanding \eqref{eq:wave0} is that of \textit{superradiance}, the fact that the conserved $\dd_t$ energy is not positive definite and thus does not control the solution $\psi$. After an appropriate frequency localisation in the frequency parameters $\w$ and $m$ (corresponding to the Killing fields $\dd_t$ and $\dd_\phi$ respectively), the superradiant frequencies are seen to be those satisfying
\be 
\label{eq:superrad}
0 \le {m\omega}\le  \frac{a m^2}{2Mr_+ - Q^2} .
\ee
 In particular, the $\dd_t$ energy identity does not preclude finite-energy exponentially growing mode solutions (with explicit $t$, $\phi$ dependence $e^{-i \w t} e^{i m \phi}$  ) associated with the frequencies \eqref{eq:superrad}, with $\w$ in the upper half-plane. The statement that such modes do not exist is known as \textit{mode stability}. In the Kerr case, this has indeed been proven  by Whiting in the celebrated \cite{Wh89}.
 
The proof of quantitative boundedness and decay for solutions of \eqref{eq:wave0} in the Kerr case given in \cite{DRS14} in fact depended on a quantitative refinement of Whiting's \cite{Wh89}. The necessary refinement was proved very recently by Shlapentokh-Rothman in \cite{SR} by first extending \cite{Wh89} to exclude resonances on the real axis and then refining this qualitative statement to a  quantitative estimate.\footnote{In the case $\abs{a} \ll M$, one need not appeal to Whiting's \cite{Wh89} (or its refinement \cite{SR}) as the small parameter may be exploited to deal directly with superradiance. A boundedness result had been obtained for $\abs{a} \ll M$ in \cite{DR11} followed by decay results in \cite{AnBl}, \cite{DR09-1} and \cite{TaTo}. For the situation in the extremal case $\abs a = M$, see \cite{Ar11-2} and \cite{Ar12}. For the case where $\Lambda > 0$, see \cite{Dy10} and for the $\Lambda < 0$ case, see \cite{Ga12}, \cite{HS13-1} and \cite{HS13-2}.}

Turning to the Kerr--Newman spacetimes, even the analogue of Whiting's mode stability is absent in the literature. In the present paper, we will prove for these spacetimes both the qualitative mode stability results (in the upper half-plane and on the real axis) as well as the quantitative estimate in the spirit of \cite{SR}. In particular, the latter result is needed for the general boundedness and decay results presented in the companion paper \cite{C14-1}. The precise mode stability results are stated here in \S\ref{sec:results} and the estimate needed in \cite{C14-1} is presented here in Theorem \ref{thm:ILED}.
 
In the Kerr case, the crucial ingredients in the proof of mode stability given in \cite{Wh89} and \cite{SR} are the remarkable transformation properties of the radial ODE satisfied by the modes. Miraculously, all the essential elements of this structure are preserved in passing from the Kerr to the Kerr--Newman solution. In particular, we show that the radial ODE can be represented as a confluent Heun equation (See \S\ref{sec:inhomogeneous}).  We then define the \textit{Whiting transform}  for $u(\w,m,\lambda,r)$ with $Im(\w) \ge 0$ (see \eqref{eq:WhitingTransform} for the definition). The Whiting transform takes the solution $u^*$ of a confluent Heun equation to $\tilde u$ which solves another confluent Heun equation with different coefficients (See Proposition \ref{prop:WhitingProperties}). There are three pivotal facts about this transform:

\begin{enumerate}[{(a)}]

\item The potential in the confluent Heun equation satisfied by $\tilde u$ possesses certain positivity properties. (See Proposition \ref{prop:positivity}.)

\item $\tilde u$ has `good' asymptotics near the horizon and near null infinity. (See Propositions \ref{prop:asymptUpper} and \ref{prop:asymptReal}.) 

\item For $\w \neq 0$ on the real axis, the limit of $u$ at the horizon is a positive multiple of the limit of $\tilde u$ at $r \rightarrow \infty$. (See Proposition \ref{prop:asymptReal}.)

\end{enumerate}

The statements above were shown to be true for the Kerr case in \cite{Wh89} and \cite{SR}; there is no a priori reason why one would expect these properties to carry over to the Kerr--Newman case. It is thus a fortunate fact that the potential and $\Delta$ parameter for the Kerr--Newman case differ from those in the Kerr case in such a way that the conditions (a), (b) and (c) still hold.
This is discussed further in \S\ref{sec:Whiting}.

For an introduction to many concepts relevant to this paper and an overview of the Kerr case, the reader is referred to the lecture notes \cite{DR08}, the survey paper \cite{DR12} and the recent \cite{DRS14}. For background on the  Kerr--Newman spacetimes, see  \cite{MR2014957}, which deals with the Dirac equation.

\section{Preliminaries}

\subsection{The Kerr--Newman spacetime}

A subextremal Kerr--Newman manifold describes a stationary spacetime in which there is a rotating charged black hole. The Kerr--Newman metric depends on three physical parameters: the mass $M$, angular momentum density $a$ and charge density $Q$. These parameters are expressed in ``natural units'' where the gravitational constant and speed of light have been set to unity ($G = c = 1$). 
 
Here we consider the \textit{subextremal} family of Kerr--Newman spacetimes in which a black hole is present. Subextremal means that $0 \le  a^2 + Q^2 < M^2.$
The other cases, $a^2 + Q^2 = M^2$ (extreme Kerr--Newman) and $a^2 + Q^2 > M^2$ (fast Kerr--Newman) have profoundly different structure. 

We will often drop the dependence of the metric on the parameters $(a,Q,M)$ and denote an arbitrary member of $\set{g_{a,Q,M} ~:~ a^2 + Q^2 < M^2}$ by $g$.

We set, for a fixed triplet of parameters $(a, Q,M)$,
$$
r_\pm := M \pm \sqrt{M^2 - a^2 - Q^2}
$$
and define the manifold $\mathcal M$ to be covered by a ``Boyer--Lindquist'' coordinate chart\footnote{This coordinate chart is global modulo the degeneration of polar coordinates.}
$$
\mathcal M = \set{(t, r, \theta, \phi) \in \mathbb R \times (r_+, \infty) \times \mathbb S^2}.
$$
The Kerr--Newman metric in these coordinates is 
\begin{eqnarray} \label{eq:metric}
g_{M,a,Q} 
&=& 
-\frac{\Delta}{\rho^2}\left(dt - a \sin^2\theta d\phi \right)^2 
+
\frac{\sin^2\theta}{\rho^2}\Big((r^2+a^2)d\phi - a dt\Big)^2 
+ 
\frac{\rho^2}{\Delta}dr^2 
+
 \rho^2 d\theta^2, \\
\mr{where}~ 
\Delta &=& r^2 - 2Mr + a^2 + Q^2  = (r - r_+)(r - r_-)
\nn\\
\mr{and}~ 
\rho^2 &=& r^2 + a^2 \cos^2\theta .\nn
\end{eqnarray}

It will be useful to define another coordinate $r^*(r) : (r+, \infty) \rightarrow (-\infty, \infty)$ (up to a constant) by
$$
\frac{dr^*}{dr} = \frac{r^2 + a^2}{\Delta}.
$$

The manifold $\mathcal M$ can be extended to a larger manifold $\tilde{\mathcal M}$.
The degeneration of the Boyer--Lindquist coordinates at $r = r_+$ is remedied by introducing the Kerr--Newman star coordinate chart $(t^*,r,\theta, \phi^*)$, where:
\be
\label{eq:changeCoords}
\left\{ 
\begin{array}{ll}
t^* = t + \bar t(r), & ~~ d\bar t̄(r) = \frac{r^2 + a^2}{\Delta},
\\
\mr{and} ~~\phi^* = \phi + \bar \phi(r), & ~~ d\bar \phi(r) = \frac{a}{\Delta}.
\end{array}
\right. 
\ee
From this, one sees that the metric extends smoothly to a metric $\tilde g$ (defined by the expression arising from applying \eqref{eq:changeCoords} to \eqref{eq:metric}) on
$$
\mathcal{\tilde M}  = \set{ (t^*, r, \theta, \phi^*) \in \mathbb R \times (r_-, \infty) \times \mathbb S^2}.
$$
Note that  $\Hp := \set{r = r_+} = \dd \mathcal{ M} \subset \tilde{\mathcal{M}}$ is a null hypersurface in $\tilde{\mathcal{M}}$. We shall refer to $\Hp$ as the horizon.

\subsection{Mode solutions of the wave equation}

For the Kerr--Newman metric in Boyer--Lindquist coordinates, the wave equation is 
\begin{equation}
\label{eq:waveKN}
\frac{1}{\rho^2\sin\theta}\left[\left( a^2\sin^2\theta - \frac{(a^2 + r^2)^2 }{\Delta} \right) \dt^2\psi  
- 
\frac{a^2}{\Delta}\dd_\phi^2\psi  
- 
\frac{2a (r^2 + a^2-\Delta  )}{\Delta}\dt\dd_\phi\psi 
+ 
\dd_r(\Delta \dd_r \psi) 
+ 
\Lang_{\mathbb S^2} \psi\right] 
= 0,
\end{equation} 
where $\Lang_{\mathbb S^2}$ denotes the (unit) spherical Laplacian:
$$
\Lang_{\mathbb S^2} = \frac{1}{\sin\theta} \dd_\theta(\sin\theta \dd_\theta) + \frac{1}{\sin^2\theta}\dd_\phi^2\psi.
$$
A general subextremal Kerr--Newman metric possesses only the two Killing fields $\dd_t$ and $\dd_\phi$. Nonetheless, Carter discovered in \cite{Ca68-2} that \eqref{eq:waveKN} can be formally separated. This is related to the existence of an additional hidden symmetry. We use this to make the following definition:

\begin{definition}\label{def:modeSol}
Let $(\mathcal M, g)$ be a subextremal Kerr--Newman spacetime. A smooth solution $\psi$ of the wave equation \eqref{eq:waveKN} is called a \emph{mode solution} if there exist $(\w,m,\ell) \in \mathbb C \setminus \set 0 \times  \mathbb Z \times \set{\mathbb{Z} ~:~ \ell \ge \abs{m} }$
such that
$$
\psi(t,r,\theta,\phi) = R\ind(r)  S\ind(\theta) e^{i m \phi} e^{-i \w t}, 
$$
where
\begin{enumerate}[1.]
\item $S\ind$ solves the following Sturm-Liouville problem
\be
 \ds{\frac{1}{\sin\theta} 
 \frac{d}{d\theta} 
 \left(\sin\theta \frac{dS\ind}{d\theta}\right) 
 -
 \left( \frac{m^2}{\sin^2\theta} - a^2 \w^2 \cos^2\theta  \right) 
 +
 \lambda\ind S\ind = 0}  \label{eq:angularODE} 
\ee
with the  boundary condition that 
\be 
e^{i m \phi}S\ind(\theta)~{extends~ smoothly~ to~}\mathbb S^2,
\label{eq:angularBC}
\ee
with $S\ind$ an eigenfunction with corresponding eigenvalue $\lambda\ind$ of the angular ODE \eqref{eq:angularODE}.\footnote{The Sturm--Liouville problem admits a set of eigenfunctions $\set{S\ind}_{\ell = \abs m}^\infty$ and real eigenvalues $\set{\lambda\ind}_{\ell = \abs m}^\infty$. The eigenfunctions $\set{S\ind}$ are called ``oblate spheroidal harmonics'' and define an orthonormal basis for $L^2(\sin\theta d\theta)$.}

\item $R$ solves the radial equation
\begin{eqnarray}\label{eq:Carter-r}
\left[ 
\dd_r(\Delta \dd_r)  
- 
\w^2
\left(  
a^2 - \frac{(a^2 + r^2)^2 }{\Delta} 
\right) 
+ 
\frac{a^2m^2}{\Delta}  
- 
\frac{2am \omega (2Mr - Q^2)}{\Delta}  
- \lambda\ind
\right]
R
= 0. 
\end{eqnarray}

\item $\ds{R(r)(r-r_+)^{-\frac{i(am - (2Mr_+ - Q^2)\w)}{r_+ - r_-} } }$ is smooth at $r = r_+$.\footnote{We will subsequently denote this as
$R(r) \sim  (r-r_+)^{\frac{i(am - (2Mr_+ - Q^2)\w)}{r_+ - r_-} }$ at $r = r_+$. }

\item There exist constants $\set{C_k}_{k=0}^\infty$ such that for any $N \ge 1$, 
$$
\ds{R(r^*) = \frac{e^{i\w r^*} }{r} \sum_{k = 0}^N C_k r^{-k} + O(r^{-N-2}),~}
$$
for large $r$.\footnote{We will subsequently denote this as $R(r^*) \sim  r^{-1} e^{i\w r^*}$ as $r \rightarrow \infty$. } 

\end{enumerate}

\end{definition}

The boundary conditions \eqref{eq:angularBC} and in points 3 and 4 above are uniquely determined by requiring that $\psi$ extends smoothly to the horizon $\Hp$ and has finite energy along asymptotically flat hypersurfaces for $Im(\w) > 0$ and along hyperboloidal hypersurfaces for $Im(\w) \le 0$. See the discussion in \cite[Appendix D]{SR} for details, cf. \cite{Dy10} and \cite{Wa13}.
 
It is convenient to define 
\be \label{eq:defu}
u_{m\ell}^{(a\omega)}(r^*) = \sqrt{r^2 + a^2}R_{m\ell}^{(a\omega)}(r)
\ee
which satisfies
\be \label{eq:Carter} 
\frac{d^2}{(dr^*)^2}u\ind(r^*) + \left(\omega^2 - V\ind(r)\right)u\ind = 0,
\ee 
where
\bea
V\ind(r) 
=
\frac{2am\omega (r^2 + a^2 - \Delta) 
- 
a^2 m^2 
+ 
\Delta \cdot (\lambda\ind + a^2\omega^2)}{(r^2 + a^2)^2}  
+ 
\frac{\Delta(  r^2 + \Delta + 2Mr)}{(r^2 + a^2)^3} 
- 
\frac{3 \Delta^2 r^2}{(r^2 + a^2)^4}.
\eea
Note that even though $R\ind$ is complex-valued, the potential  $V\ind$ is real.

We will often drop the indices $\w, m, \ell$ when there is no risk of confusion. We will also adopt the convention that $u'$ denotes a derivative with respect to $r^*$.

\subsection{The Wronskian}
\label{sec:Wronskian}

Through asymptotic analysis of  \eqref{eq:Carter}, one can make the following definitions:

\begin{definition}
Let $u_{hor}(r^*, \w, m, \ell)$ be the unique function satisfying 

\begin{enumerate}[1.]

\item $u_{hor}'' + (\w^2 - V) u_{hor} = 0$.

\item $u_{hor} \sim  (r-r_+)^{\frac{i(am - (2Mr_+ - Q^2)\w)}{r_+ - r_-} }$ as $r^* \rightarrow -\infty$.

\item $\ds{ \abs{ \left( (r(r^*)-r_+)^{-\frac{i(am - (2Mr_+ - Q^2)\w)}{r_+ - r_-} } u_{hor} \right) (-\infty)}^2 = 1 }$.

\end{enumerate}

\end{definition}

\begin{definition}
Let $u_{out} (r^*, \w, m, \ell )$ be the unique function satisfying 

\begin{enumerate}[1.]

\item $u_{out}'' + (\w^2 - V) u_{out} = 0$.

\item $u_{out} \sim  e^{i\w r^*}$ as $r^* \rightarrow \infty$.

\item $ \abs{ \left(   u_{out} e^{-i\w r^*} \right) (\infty)  }^2 = 1$.

\end{enumerate}
\end{definition}

One then defines the Wronskian
\be \label{eq:W}
W(\w,m,\ell ) = u_{hor}(r^*) u_{out}'(r^*) - u_{hor}'(r^*) u_{out}(r^*) .
\ee
The Wronskian can be evaluated at any fixed $r^*$. The Wronskian $W$ will vanish if and only if the solutions are linearly dependent. Then $W = 0$ implies $\abs{W^{-1}} = \infty$. The quantitative mode stability result will be an explicit upper bound for  $\abs{W^{-1}}$, so that $u_{out}$ and $u_{hor}$ are linearly independent and any solution of the Carter ODE \eqref{eq:Carter} can be expressed as a superposition of those solutions.

\subsection{The inhomogeneous equation}
\label{sec:inhomogeneous}

In the proof of Theorem \ref{thm:quantModeStab}, we will consider the following inhomogeneous form of \eqref{eq:Carter-r},
\be 
\left[ 
\dd_r(\Delta \dd_r)  
- 
\w^2
\left(  
a^2 - \frac{(a^2 + r^2)^2 }{\Delta} 
\right) 
+ 
\frac{a^2m^2}{\Delta}  
- 
\frac{2am \omega (2Mr - Q^2)}{\Delta}  
- \lambda\ind
\right]
R_{m\ell}^{(a\omega)} 
= F ,
\label{eq:CarterF} 
\ee
where  $F$ is a compactly supported smooth function on $(r_+,\infty)$.
The corresponding inhomogeneous version of \eqref{eq:Carter} is then
\be 
\frac{d^2}{(dr^*)^2}u\ind(r^*) 
+ 
\left(\omega^2 - V\ind(r)\right)u\ind 
=
H := \frac{\Delta F}{(r^2 + a^2)^{1/2}}.
\label{eq:CarterH} 
\ee

\section{Statement of mode stability results}
\label{sec:results}

For a subextremal Kerr--Newman spacetime $(\mathcal M, g)$, we have the following results.

\bt[Quantitative mode stability on the real axis]\label{thm:quantModeStab}
 Let 
 $$
 \mathcal F \subset \set{(\w, m, \ell) \in \mathbb R \times\set{ \mathbb Z \times \mathbb{Z} ~|~ \ell \ge \abs{m} }}
 $$
be a frequency range for which
$$
C_{\mathcal F} 
:= 
\sup_{(\w,m,\ell) \in \mathcal F} 
\left(\abs{\w} + \abs{\w}^{-1} + \abs{m} + \abs{\lambda\ind }\right) 
< \infty.
$$
Then the Wronskian $W$ given by \eqref{eq:W} satisfies
$$
\sup_{(\w,m,\ell) \in \mathcal F} \abs{W^{-1}}  
\le 
G(C_{\mathcal F}, a, Q , M).
$$
where the function $G$ can, in principle, be given explicitly. 
\et 

In proving the quantitative result above, we will also obtain the following qualitative results.

\bt[Mode Stability on the real axis] 
\label{thm:modeStabReal} 
There exist no non-trivial mode solutions corresponding to $ \w \in \mathbb R \setminus \set{0}$.
\et

\bt[Mode Stability] 
\label{thm:modeStabUpper} 
There exist no non-trivial mode solutions corresponding to $Im (\w) > 0$.
\et

Theorem \ref{thm:modeStabUpper} is the analogue of Whiting's original mode stability result \cite{Wh89}.  Theorem \ref{thm:modeStabReal} is the analogue of Shlapentokh-Rothman's extension of Whiting's mode stability result \cite{Wh89} to the real axis. Theorem \ref{thm:quantModeStab} is the quantitative refinement of Theorem \ref{thm:modeStabReal} needed in the companion paper \cite{C14-1} for the proof of linear stability of subextremal Kerr--Newman black holes.

Note that for non-superradiant frequencies $\w$, $m$, i.e. those outside of the range \eqref{eq:superrad}, Theorem \ref{thm:modeStabReal} and Theorem \ref{thm:modeStabUpper} follow immediately from the energy identity (see \cite[\S 1.5 \& \S 1.6]{SR}). In what follows, we will not however make a distinction between superradiant and non-superradiant frequencies.
 
\section{The Whiting transform}
\label{sec:Whiting}

The problem with trying to derive energy estimates for the Carter ODE \eqref{eq:Carter} is that the boundary condition at $r^* = -\infty$ may give a non-positive term due to superradiance. To deal with this, we will first cast \eqref{eq:Carter} as a confluent Heun equation \eqref{eq:carterHeun}. Applying the Whiting transform \eqref{eq:WhitingTransform} to \eqref{eq:carterHeun}, we will  obtain a new confluent Heun equation \eqref{eq:carterWhiting} with different coefficients and boundary conditions that allow for a useful energy estimate.

\subsection{The confluent Heun equation}
\label{sec:CHE}
We rescale $R$ as follows. Let 
\be
u^* := e^{{i \w r}}
(r- r_-)^{-\eta}
(r- r_+)^{-\xi}
R(r)
\ee
where
$$
\eta := -\frac{i \left(a m-\omega  \left(2 M {r_-}-Q^2\right)\right)}{{r_+}-{r_-}}
~~~~and ~~~~ 
\xi := \frac{i \left(a m-\omega  \left(2 M {r_+}-Q^2\right)\right)}{{r_+}-{r_-}}.
$$
Then $g$ satisfies the following Confluent Heun equation: 
\be \label{eq:carterHeun}
(r-r_+)(r-r_-)\frac{d^2 u^*}{dr^2}
+
\left(
\gamma (r - r_+)
+
\delta (r- r_-)
+ 
p(r-r_+)(r-r_-)
\right)
\frac{du^*}{dr}
+
\left(
\alpha p  (r- r_-)
+ 
\sigma
\right) u^* = G
\ee
where
\bea
\gamma &:=& 2\eta + 1,
\\
\delta &:=& 2\xi + 1,
\\
p &:=&-2 i \w,
\\
\alpha &:=&1,
\\
\sigma &:=& 2 a m \w - 2 \w r_- i - \lambda\ind - a^2 \w^2
\\
\mr{and~}
G &:=& e^{{i \w r}}
(r- r_-)^{-\eta}
(r- r_+)^{-\xi}
F .
\eea
This can be verified by a direct calculation, generalising the analogous computation in \cite{Wh89}.

Note that, as in the (subextremal) Kerr case, $r_+$ and $r_-$ are distinct roots of $\Delta$. If $\Delta$ had more roots, or if these roots were not distinct, the Carter ODE would lie in a different class of equations. 

\subsection{The transformed equation}

We now generalise the Whiting transformation to the Kerr--Newman case.

\begin{proposition}
\label{prop:WhitingProperties}
Let $Im(\w) \ge 0$, $\w \ne 0$, and let $R$ solve \eqref{eq:CarterF} with the boundary conditions of Definition \ref{def:modeSol}. Define $\tilde u$ by 
\be  \label{eq:WhitingTransform}
\tilde u(x^*) 
:= 
(x^2 + a^2)^{1/2}
(x- r_+)^{-2iM\w}
e^{-i\w x} 
\int_{r_+}^\infty 
e^{\frac{2 i \w}{r_+ - r_- } (x- r_-)(r - r_-)}
(r- r_-)^\eta
(r- r_+)^\xi
e^{-i\w r}
R(r) dr
\ee
where
$$
\eta := -\frac{i \left(a m-\omega  \left(2 M {r_-}-Q^2\right)\right)}{{r_+}-{r_-}}
~~~~and ~~~~ 
\xi := \frac{i \left(a m-\omega  \left(2 M {r_+}-Q^2\right)\right)}{{r_+}-{r_-}}.
$$
Then $\tilde u(x)$ is smooth on $(r_+,\infty)$ and satisfies the following confluent Heun equation:
\be \label{eq:carterWhiting}
\tilde u'' + \Phi \tilde u = \tilde H, 
\ee
where primes denote derivatives with respect to  $x^*$ (and $\frac{dx^*}{dx} = \frac{x^2 + a^2}{\Delta}$),
\bea
\tilde H(x^*) &:=&\frac{ (x- r_+)(r - r_-)}{(x^2 + a^2)^2} \tilde G(x),
\\
\tilde G(x) &:=& 
\frac{(x^2 + a^2)^{1/2}}{
(x- r_+)^{2iM\w}}
e^{-i\w x} 
\int_{r_+}^\infty 
e^{\frac{2 i \w}{r_+ - r_- } (x- r_-)(r - r_-)}
(r- r_-)^\eta
(r- r_+)^\xi
e^{-i\w r}
F(r) dr
\\
and ~~ \Phi(x^*) 
&:=&
\frac{(x-r_-) (x-r_+) }{\left(a^2+x^2\right)^4}
\left(
\left(2x^2 -a^2\right) ( r_- r_+)-2 M x (x^2 - 2a^2) -3 a^2x^2
\right)
\\
&&
+
\frac{(x-r_-) (x-r_+) }{\left(a^2+x^2\right)^2}
\left(
\frac{4 a m (x-M) \omega }{r_--r_+}
-
\lambda\ind  -a^2 \omega ^2 \right.
\\
&&
\left.
+
\frac{8 M^2 (x-M) (x-r_-) \omega ^2}{(r_--r_+) (r_+-x)}
+
\frac{(x-r_-) \left((r_+-r_-) (x-r_+)-4 Q^2\right) \omega ^2}{r_+-r_-}\right)
\eea

\end{proposition}

\bp
It turns out that the proof is a direct modification of the computations in \cite[\S 4]{SR}. Let us remark on the fortuitous structure of the Kerr--Newman spacetimes that makes this so. We have already remarked in \S\ref{sec:CHE} that \eqref{eq:carterHeun} is a confluent Heun equation and thus (at least formally) admits non-trivial transformations. The exponents $\eta$ and $\xi$ are obtained from indicial equation associated to \eqref{eq:Carter}. They are the unique exponents that give the correct asymptotics at $r_+$ and $r_-$. 

The definitions of $\eta$, $\xi$, $r_+$ and $r_-$  for the Kerr--Newman case differ from those in the Kerr case, but the potential $V\ind$, the parameter $\Delta$ and  the asymptotics of the solutions of mode solutions of \eqref{eq:Carter}, have the same structure. The convergence of the integral in \eqref{eq:WhitingTransform} thus follows as in \cite[\S 4]{SR}.
\ep

\noindent
\textbf{Remark.} The Whiting transform is a shifted, rescaled Fourier transform of a rescaled version of $R$. This fact will be crucial in showing that the vanishing of $\tilde u$ forces $R$ to vanish.

\subsection{Asymptotics of the transformed solution}

The good asymptotic properties of $\tilde u$ (c.f.~(b) and (c) of the introduction) are encapsulated in the following two propositions.

\begin{proposition}
\label{prop:asymptUpper}
Let $\w$ and $\tilde u$ be as in the statement Proposition \ref{prop:WhitingProperties}. If $Im(\w) > 0$ then 

\begin{enumerate}[1.]

\item $\tilde u = O \left( (x - r_+)^{2M Im(\w) } \right)$ as $x \rightarrow r_+$.

\item $\tilde u' = O\left( (x - r_+)^{2M Im(\w) } \right)$ as $x \rightarrow r_+$.

\item $\tilde u = O\left( e^{- Im(\w) x^{1 + 2M Im(\w) }} \right)$ as $x \rightarrow \infty$.

\item $\tilde u' = O\left( e^{- Im(\w) x^{1 + 2M Im(\w) }} \right)$ as $x \rightarrow \infty$.
\end{enumerate}

\end{proposition}

\begin{proposition}
\label{prop:asymptReal}
Let $\w$ and $\tilde u$ be as in the statement Proposition \ref{prop:WhitingProperties}. If $\w \in \mathbb{R} \setminus \set 0$ then 

\begin{enumerate}[1.]

\item $\tilde u$ and  $\tilde u'$ are uniformly bounded.

\item 
$\abs{\tilde u(\infty)}^2 
= 
\frac{(r_+ - r_-)^2 \abs{\Gamma(2\xi + 1)}^2 }{4(2 M  r_+ - Q^2)\w^2 } 
\abs{u(-\infty)}^2$, where $\Gamma(z) := \int_0^\infty e^{-t} t^{z-1} dt$ is the Gamma function.

\item $\tilde u' - i \w \tilde u = O\left( x^{-1} \right)$ as $x^* \rightarrow \infty$.

\item $\tilde u'  + i \w r_+^{-1}(r_+ - r_-) \tilde u= O\left( x - r_+ \right)$ as $x \rightarrow -\infty$.
\end{enumerate}
\end{proposition}

The proofs of these propositions are direct modifications of the computations in \cite[\S 4]{SR}.

For all the results above, except Proposition \ref{prop:asymptReal}.2, the difference between the Kerr and Kerr--Newman case is encapsulated within the different definitions of $r_+$ and $r_-$. 

Proposition \ref{prop:asymptReal}.2 is exceptional in that we see an explicit difference from the Kerr case. This is due to the presence of $(2 M  r_+ - Q^2)$ in the null generator of the Kerr--Newman horizon. 

Proposition \ref{prop:asymptReal}.2 is crucial in proving the quantitative result Theorem \ref{thm:quantModeStab} as it provides a correspondence between the horizon asymptotics of the solution of the Carter ODE and the large $r^*$ asymptotics of the transformed solution. This correspondence is what allows for the quantitative estimate of the horizon flux in terms of the inhomogeneity $F$ (see the proof of Proposition \ref{prop:WronskianEst}).

We can now prove the qualitative Theorems \ref{thm:modeStabReal} and  \ref{thm:modeStabUpper}.

\section{Proofs of mode stability}
\label{sec:proofs}

\subsection{Qualitative results}
\label{sec:qualitative}

The final element of the structure necessary to prove mode stability for the Kerr--Newman spacetimes is the following positivity property (c.f.~(a) of the introduction):

\begin{proposition}
\label{prop:positivity}
Under the conditions of Proposition \ref{prop:WhitingProperties}, 
$$
Im(\Phi \bar \w) \ge 0.
$$
If $\w \in \mathbb R \setminus \set 0$, then $\Phi$ is real-valued.
\end{proposition}

\bp
The second statement is clear from the definition of $\Phi$.
A (tedious) computation shows that
 \bea
 Im(\Phi \bar \w)
  &=&
  \frac{(x-r_-) (x-r_+) }{\left(a^2+x^2\right)^2}  
  Im\left((-\lambda\ind  -a^2\w^2) \bar \w\right)
    +
  \frac{(x-r_-)^2 (x-r_+)^2 \w_I \abs{\w}^2 }{
  \left(a^2+x^2\right)^2} 
  \\
  && 
  +
    \frac{(x-r_-) (x-r_+) }{\left(a^2+x^2\right)^4}
  {(\w_I)  
  \left(x^2(r_+ - a^2 - Q^2) + r_-(x^2 + a^2)(x-r_+) + 2xa^2(x + r_- - r_+)\right)}
\\
&&
+  
  \frac{(x-r_-) (x-r_+) \w_I \abs{\w}^2 }{
  \left(a^2+x^2\right)^2} 
\frac{(x-r_-) \left(8 M^2 (x-M)-4 Q^2 (x-r_+)+(r_+-r_-) (x-r_+)^2\right)}
{(r_+-r_-) (x-r_+)}.
\eea
To see that $Im\left((-\lambda\ind  -a^2\w^2) \bar \w\right) \ge 0$, multiply \eqref{eq:angularODE} by $\overline{ \w S\ind}$ and integrate by parts over $[0,\pi]$. 
 
The positivity of the other terms follows from the following chain of inequalities
$$
0 \le r_- \le M \le r_+ \le x 
$$
and the subextremal condition $a^2 + Q^2 < M^2$.
\ep

We define the \textit{microlocal energy current}
$$
\tilde Q_T := Im(\tilde u' \overline{\w \tilde u}).
$$

 \begin{proof}[Proof of Theorem \ref{thm:modeStabUpper} (Mode stability in the upper half-plane)]
 
 Let $\w = \w_R + i\w_I$ and $Im(\w) = \w_I > 0$ and consider a mode solution of \eqref{eq:waveKN} with $(u\ind, S\ind, \lambda\ind)$.
Define $\tilde u$ to be the \eqref{eq:WhitingTransform} of $u\ind$. Then Proposition \ref{prop:asymptUpper} implies that $\tilde Q_T(\pm \infty) = 0$ so
$$
0 
= 
-\int_{-\infty}^\infty (\tilde Q_T)' dr^* 
=  
\int_{-\infty}^\infty \w_I \abs{\tilde u'}^2 + Im(\Phi \bar \w) \abs{\tilde u}^2  dr^*
$$
Since Proposition \ref{prop:positivity} guarantees that $Im(\Phi \bar \w) \ge 0$, we conclude that, $\tilde u$, the Whiting transform of $u$ vanishes. Hence
$$
\tilde R(x) := \int_{r_+}^\infty 
e^{\frac{2 i \w}{r_+ - r_- } (x- r_-)(r - r_-)}
(r- r_-)^\eta
(r- r_+)^\xi
e^{-i\w r}
R(r) dr = 0.
$$
Extending $R$ by $0$, we see that the Fourier transform of $(r- r_-)^\eta
(r- r_+)^\xi
e^{-i\w r}
R(r)$ is (up to a change of variable)
$$
\hat R(z) := \int_{-\infty}^\infty 
e^{{2 i \abs{\w}^2}(z- r_-)}
(r- r_-)^\eta
(r- r_+)^\xi
e^{-i\w r}
R(r) dr .
$$
Since $R$ is supported in $[0,\infty)$, $\hat R$  can be extended holomorphically into the upper half plane.
Since $\tilde R$ vanishes on $x \in (r_+, \infty)$, $\hat R = 0$ on the line $\set{z = \bar \w (x - r_+)/(r_+ - r_-)~| ~ x\in (r_+,\infty)}$. Analyticity then implies that $\hat R$ and hence $R$ vanish everywhere. 
\end{proof}

\begin{lemma}[Unique continuation \cite{SR}]
\label{lemma:UniqueContinuation} Suppose that we have a solution $u(r^*) : (-\infty,\infty) \rightarrow \mathbb C$ to 
$$
u + (\w^2 - V )u = 0
$$
such that
\begin{enumerate}[1.]
\item $\w \in \mathbb R \setminus \set 0$,
\item $u$ is uniformly bounded and $(\audr + \au)(\infty) = 0$,
\item  $V$ is real, uniformly bounded, $V = O(r^{-1})$ as $r \rightarrow \infty$  and $V' = O(r^{-2})$ as $r \rightarrow \infty$.
\end{enumerate}
Then $u$ is identically $0$.
\end{lemma}

\bp
This follows exactly as in \cite[\S 6]{SR}
\ep

\begin{proof}[Proof of Theorem \ref{thm:modeStabReal} (Mode stability on the real axis)]

 Let $\w \in \mathbb R \setminus \set 0$ and consider a mode solution of \eqref{eq:waveKN} with $(u\ind, S\ind, \lambda\ind)$.
Define $\tilde u$ by \eqref{eq:WhitingTransform}. By Proposition \ref{prop:WhitingProperties}, $\Phi$ is real, so  
$(\tilde Q_T)' = 0$ . Hence $\tilde Q_T( \infty) - Q_T( -\infty)  = 0$.
The boundary conditions from Proposition \ref{prop:asymptReal} then imply that
$$
\w^2\abs{\tilde u(\infty)}^2
+
\abs{\tilde u'(\infty)}^2
+
\w^2 \frac{r_+ - r_-}{r_+} \abs{\tilde u(-\infty)}^2
+
\frac{r_+ }{r_+- r_-} \abs{\tilde u'(-\infty)}^2 = 0.
$$
By Lemma \ref{lemma:UniqueContinuation}, we conclude that $\tilde u$ vanishes. 

Extending $R$ by $0$, we see that 
$$
\tilde R(y) := \int_{-\infty}^\infty 
e^{\frac{2 i \w}{r_+ - r_- } (y- r_-)(r - r_-)}
(r- r_-)^\eta
(r- r_+)^\xi
e^{-i\w r}
R(r) dr 
$$
vanishes for $\set{y \in (r_+,\infty)}$. Since the Fourier transform of a non-trivial function supported in $(0, \infty)$ cannot vanish on an open set, $R$ must vanish everywhere.  
\end{proof}

\subsection{Quantitative results}
\label{sec:quantitative}

The strategy is to express $\tilde u$ in terms of the functions $u_{out}$ and $u_{hor}$ and $W$ defined in \S\ref{sec:Wronskian} and obtain an an estimate for $W^{-1}$ in terms of $u(-\infty)$. This quantity is then estimated using the ODE \eqref{eq:CarterH}. 

\begin{proposition}
\label{prop:WronskianEst}
Define $\mathcal F$ as  in Theorem \ref{thm:quantModeStab}. For  $(\w,m,\ell) \in \mathcal F$  let  $u$ solve \eqref{eq:CarterH} with $H(x^*)$ a smooth, compactly supported function. Then for $\epsilon > 0$, there exists a positive constant $C := C({\mathcal F}, a, Q , M)$ such that
 $$
 \abs{u(-\infty)}^2 
 \le 
 C \left( 
 \epsilon^{-1}\int_{r_+}^\infty \abs{F(r)}^2r^4 dr
\right).
 $$
\end{proposition}

\bp
Since  $(\tilde Q_T)' = \w Im(\tilde H \bar u)$, 
$$ 
\int_{-\infty}^\infty \w Im(\tilde H \bar u)dr^* 
=  
\tilde Q_T(\infty) - \tilde Q_T(-\infty). 
$$
The boundary conditions from Proposition \ref{prop:asymptReal} imply that
$$
\w^2\abs{\tilde u(\infty)}^2
+
\abs{\tilde u'(\infty)}^2
+
\w^2 \frac{r_+ - r_-}{r_+} \abs{\tilde u(-\infty)}^2
+
\frac{r_+ }{r_+- r_-} \abs{\tilde u'(-\infty)}^2
= \int_{-\infty}^\infty \w Im(\tilde H \bar u)dr^* .
$$
So changing variables,  applying the Plancherel identity and the Cauchy Schwarz inequality, we have  
$$
\w^2 \abs{\tilde u(\infty)}^2 
\le 
\int_{-\infty}^\infty \w Im(\tilde H \bar u)dr^*
 \le 
 C \left( 
 \epsilon^{-1}\int_{r_+}^\infty \abs{F(r)}^2r^4 dr
 +
\epsilon  \int_{r_+}^\infty \abs{R(r)}^2 dr
\right).
 $$
 Then by Proposition \ref{prop:asymptReal} 
 $$
\abs{ u(-\infty)}^2 
= 
\frac{4 \w^2(2M r_+ - Q^2)}{\abs{\Gamma(2\xi + 1)}^2 } \abs{\tilde u(\infty)}^2 
\le 
 C \left( 
 \epsilon^{-1}\int_{r_+}^\infty \abs{F(r)}^2r^4 dr
 +
\epsilon  \int_{r_+}^\infty \abs{R(r)}^2 dr
\right).
 $$
 Finally, 
 $$
  \int_{r_+}^\infty \abs{R(r)}^2 dr
\le 
C \int_{r_+}^\infty \abs{F(r)}^2r^4 dr,
 $$
 by the same argument as found in \cite[\S 5]{SR}. 
\ep

For the quantitative result, we construct mode solutions solutions to the Carter ODE from the Wronskian and apply the proposition above.

\begin{lemma}
\label{lemma:inhomContruction}
Let $H(x^*)$ be compactly supported. For any $(\w,m,\ell) \in \mathcal F$ (where $\mathcal F$ is as defined in Theorem \ref{thm:quantModeStab}), the function
$$
u({r^*}) = W(\w,m,\ell)^{-1}
\left( 
u_{out}(r^*)  
\int_{-\infty}^{r^*} 
u_{hor}(x^*) H(x^*) dx^*
+
 u_{hor}(r^*)
 \int^{\infty}_{r^*} 
  u_{out}(x^*) H(x^*) 
  dx^*
 \right)
$$
satisfies
$$
u'' + (\w^2 - V) u = H
$$
and the boundary conditions of a mode solution (see Definition \ref{def:modeSol}).
\end{lemma}

\bp
This is verified by a direct calculation.
\ep 

 \begin{proof}[Proof of Theorem \ref{thm:quantModeStab} (Quantitative mode stability on the real axis)]
Define $\tilde u$ by Lemma \ref{lemma:inhomContruction}. Then
 $$
 \abs{u(-\infty)}^2 
=
 \abs{W^{-2}} \abs{\int^{\infty}_{-\infty} 
  u_{out}(x^*) H(x^*) 
  dx^* }^2.
 $$
 Rearranging this expression and applying Proposition \ref{prop:WronskianEst} we find that
 $$
 \abs{W^{-2}} 
=
\frac{ \abs{u(-\infty)}^2 }
{ \abs{\int^{\infty}_{-\infty} 
  u_{out}(x^*) H(x^*) 
  dx^* }^2 } 
 \le
 C
  \frac{\int^{\infty}_{-\infty} \abs{
  (r^2 + a^2)^{1/2} \Delta^{-1} H(x^*) } r^4
  dr  }
  { \abs{\int^{\infty}_{-\infty} 
  u_{out}(x^*) H(x^*) 
  dx^* }^2 } .
 $$
 Note that by Proposition \ref{prop:asymptReal}, for sufficiently large $x$, $\abs{u_{out}(x) - e^{i\w x} }< Cx^{-1}$ for an an explicit $C$. Since $W$ is independent of $H$ we choose a compactly supported $H$ for which the right hand side of the estimate above is finite. We thus have a quantitative estimate for $\abs{W^{-2}} $.
 \end{proof}

\section{Application: Integrated local energy decay}

We now apply Theorem \ref{thm:quantModeStab} to prove Theorem \ref{thm:ILED}, which provides a quantitative energy decay estimate for solutions of the wave equation \eqref{eq:waveKN} on subextremal Kerr--Newman spacetimes which are supported in a compact range of superradiant frequencies. This is the estimate appealed to in \cite{C14-1} to control the horizon term $|{u\ind(-\infty)}|^2$ in the bounded superradiant frequency region.

We wish to apply Carter's separation to the solution of \eqref{eq:waveKN}. In order to perform this separation, we must be able to take the Fourier transform in time. We therefore deal with solutions of \eqref{eq:waveKN} which belong to the following class of functions.

\begin{definition}
A smooth function $f(t,r,\theta,\phi)$ is said to be admissible if for any multi-indices $\alpha$, $\beta$ s.t. $\abs \alpha \ge 1$, $\abs \beta \ge 0$,  we have
\begin{enumerate}[1.]
\item 
$\ds{ 
\int_{r > r_0} 
\int_{\mathbb S^2}  
\abs{\dd^\alpha f}^2
|_{t=0} r^2 ~
\sin\theta dr~ d\theta ~d\phi 
< \infty     }$ 
for sufficiently large $r_0$.

\item 
$\ds{ 
\int_{0}^\infty    
\abs{\dd^\beta f}^2
dt
< \infty }$ 
for  any $(r,\theta, \phi) \in (r_+, \infty) \times \mathbb S^2$.

\item 
$\ds{ 
\int_{0}^\infty    
\int_{K} 
\abs{\dd^\beta f}^2~
\sin\theta ~dr ~d\theta~ d\phi ~dt
< \infty }$ 
for any compact $K \in (r_+, \infty) \times \mathbb S^2$.

\end{enumerate}

For an admissible function $f$ we also define
\be \label{eq:energy}
\abs{\dd f}^2 := \abs{(\dd_t + \dd_{r^*})f}^2
+
\Delta\abs{(\dd_t - \dd_{r^*})f}^2
+
r^{-2}\left(\sin^{-2}\theta\abs{\dd_\phi f}^2 + \abs{\dd_\theta f}^2 \right).\footnote{The apparent degeneration of this energy as $r \rightarrow \infty$ is due to the hyperboloidal nature of $\Sigma_0$. The term $\Delta\abs{(\dd_t - \dd_{r^*})f}^2$ converges to the transversal derivative at the horizon.}
\ee

\end{definition}

The main application of Theorem \ref{thm:quantModeStab} in \cite{C14-1} is to admissible solutions $\psi$ of \eqref{eq:waveKN} which are cut off as follows.

\begin{definition}
Let $\Sigma_0$ be a spacelike hyperboloidal hypersurface connecting the horizon $\Hp$ and future null infinity. Let $\Sigma_1$ be the time 1 image of $\Sigma_0$ under the flow generated by $\dd_t$. Then define a smooth cut-off $\gamma$ which is identically $0$ in the past of $\Sigma_0$ and identically $1$ in the future of $\Sigma_1$.
We define $\psi_\cut := \gamma \psi$, which satisfies the inhomogeneous wave equation
\be \label{eq:cutOffWaveEq}
\square_g \psi_{\cut} = F, 
~~~{where}~~~F = (\square \gamma)\psi + 2 \nabla^\mu\gamma\nabla_\mu \psi.
\ee 
\end{definition}

\begin{proposition}[Carter's separation]
Admissible solutions $f$ of \eqref{eq:waveKN} and \eqref{eq:cutOffWaveEq} can be expressed as
\be \label{eq:decomposition} 
f(t, r, \theta, \phi) = \overbrace{\frac{1}{\sqrt{2\pi}}\int_{-\infty}^\infty \underbrace{ \sum_{m, \ell > \abs{m}} R\ind(r) \cdot S^{(a\omega)}_{m\ell}(\cos\theta) e^{i m \phi} }_{\mr{Oblate ~spheroidal ~expansion}} e^{-i\omega t} }^{\mr{Fourier ~transform}} ~d\omega.
\ee
The function $R\ind$ corresponding to $f = \psi$ solves \eqref{eq:Carter-r}.
The function $R\ind$ corresponding to $f = \psi_\cut$ satisfies the inhomogeneous equation \eqref{eq:CarterF} with $F = F\ind$, the Fourier transform of $F$ projected to the oblate spheroidal harmonic corresponding to $\lambda\ind$. The rescaled function $u\ind$ satisfies \eqref{eq:CarterH} with $H = H\ind := \Delta (r^2 + a^2)^{-1/2} F\ind$, where this equality is to be understood in the sense of $L^2_{\w \in \mathcal B }\ell^2_{ m, \ell \in \mathcal C}$. Note moreover that this $H$ is not compactly supported.

\end{proposition}

\bp
See \cite[\S 5]{C14-1}.
\ep 

\bt 
\label{thm:ILED}
Let $\psi_\cut$ be an admissible solution of \eqref{eq:cutOffWaveEq} and let
 $\mathcal B \subset \mathbb R$ 
and
 $$
 \mathcal C \subset \set{( m, \ell) \in  \mathbb Z \times \mathbb{Z} ~|~ \ell \ge \abs{m} }
 $$
  such that
$$
C_{\mathcal B} 
:= 
\sup_{\w \in \mathcal B} 
\left(\abs{\w} + \abs{\w}^{-1} \right) 
< \infty
~\mr{and}~~
C_{\mathcal C} 
:= 
\sup_{m,\ell \in \mathcal C} 
\left( \abs{m} + \abs{\lambda\ind}\right) 
< \infty.
$$
There exists a constant $K := K(r_0,r_1,  C_{\mathcal{B}}, C_{\mathcal{C}}, a ,Q, M)$ such that
\be \label{eq:ILED}
 \int_\mathcal{B} 
 \sum_{m,\ell \in \mathcal C}  
 \left(
 \left(
\abs{u\ind(-\infty)}^2 + \abs{u\ind(\infty)}^2
 \right)
 +
\int_{r_0}^{r_1} \abs{\dd_{r^*}u\ind}^2 + \abs{u\ind}^2  ~ dr^*
\right)
~d\w
 \le 
 \int_{\Sigma_0} \abs{\dd\psi}^2 ,
\ee
 where $\abs{\dd\psi}^2$  is defined by \eqref{eq:energy}, $u\ind =\sqrt{r^2 + a^2} R\ind$ and each $R\ind$ solves \eqref{eq:CarterF} for $\w  \in \mathcal B$ and $(m,\ell) \in \mathcal C$. 
\et 

\bp
For $u$ satisfying the hypotheses of the theorem, we have for any $r^* \in (-\infty, \infty)$, 
\begin{eqnarray}
\label{eq:representation}
u({r^*}) = W(\w,m,\ell )^{-1}
\left( 
u_{out}(r^*)  
\int_{-\infty}^{r^*} 
u_{hor}(x^*) H(x^*) dx^*
+
 u_{hor}(r^*)
 \int^{\infty}_{r^*} 
  u_{out}(x^*) H(x^*) 
  dx^*
 \right),
 \\
 \label{eq:drepresentation}
u'({r^*}) = W(\w,m,\ell )^{-1}
\left( 
u_{out}'(r^*)  
\int_{-\infty}^{r^*} 
u_{hor}(x^*) H(x^*) dx^*
+
 u_{hor}'(r^*)
 \int^{\infty}_{r^*} 
  u_{out}(x^*) H(x^*) 
  dx^*
 \right),
\end{eqnarray}  
where the inequalities above hold in the sense of $L^2_{\w \in \mathcal B }\ell^2_{ m, \ell \in \mathcal C}$ (see \cite[\S 3]{SR} for the full derivation of this representation).\footnote{Roughly speaking, this is the converse of Lemma \ref{lemma:inhomContruction}.} 

By the construction of $u_{hor}$ and $u_{out}$, there exists a positive $K := K( C_{\mathcal{B}}, C_{\mathcal{C}}, a ,Q, M)$ such that
\be \label{eq:uHorOutBdd}
\sup_{r^* \in \mathbb R, \w \in \mathcal B,  (m, \ell) \in \mathcal C} \left(\abs{u_{hor}} +\abs{u_{out}} \right)
< K < \infty,
\ee

Evaluating \eqref{eq:representation} at $r^* = -\infty$ and taking \eqref{eq:uHorOutBdd} into account, 
\be \label{eq:horizonControl}
\int_\mathcal{B} 
 \sum_{m,\ell \in \mathcal C}  
\abs{u\ind(-\infty)}^2 
~d\w
\le 
K
\limsup_{r^* \rightarrow \infty}
\int_\mathcal{B} 
 \sum_{m,\ell \in \mathcal C}  
 W^{-2}
 \abs{
 \int^{\infty}_{r^*} 
  u_{out}(x^*) H\ind(x^*) 
  dx^*
  }^2 
  ~ d\w.
\ee

For the term $\abs{u(\infty)}^2$ we apply the microlocal energy current:
\bea
\w^2 \abs{u\ind(\infty)}^2 
&=&
Q_T(\infty) 
=
 Q_T(-\infty) 
 + 
 \int_{-\infty}^\infty (Q_T)' dr^* 
 \\
&=&
\w(am - (2Mr_+ - Q^2)\w) \abs{u\ind(-\infty)}^2
+ 
 \w \int_{-\infty}^\infty Im(H\ind \bar u\ind) dr^* 
\eea
So by \eqref{eq:horizonControl},

\begin{eqnarray}
\label{eq:infinityControl}
\int_\mathcal{B} 
 \sum_{m,\ell \in \mathcal C}  
\abs{u\ind(\infty)}^2 
~d\w
&\le& 
K
\int_\mathcal{B} 
 \sum_{m,\ell \in \mathcal C}  
 W^{-2}
 \abs{
 \int^{\infty}_{-\infty} 
  u_{out}(x^*) H\ind(x^*) 
  dx^*
  }^2 
    ~ d\w
    \nn
  \\
  && +
 \int_\mathcal{B} 
 \sum_{m,\ell \in \mathcal C}  
  \w \int_{-\infty}^\infty Im(H\ind \bar u\ind) dr^* 
  ~ d\w.
\end{eqnarray}

For the integral term, we begin by taking $R_1$ much larger than $r_1$ and applying \eqref{eq:representation}:

\bea
\int_\mathcal{B} 
 \sum_{m,\ell \in \mathcal C}  
\sup _{r^* \in (r_0,r_1)}\abs{u\ind}^2 
~d\w
&\le& 
K
\int_\mathcal{B} 
 \sum_{m,\ell \in \mathcal C}  
 W^{-2}
\left( 
\sup _{r^* \in [r_0,r_1]} 
\abs{
\int_{-\infty}^{r^*} 
u_{hor}(x^*) H\ind(x^*) dx^*
}^2 
\right.
\nn \\
&& \left.
+
\sup _{r^* \in [r_0,r_1]} 
\abs{
 \int_{r^*}^{R_1} 
  u_{out}(x^*) H\ind(x^*) 
  dx^*
 }^2
\right.
\nn \\
&& \left.
+
\abs{
 \int_{R_1} ^{\infty}
  u_{out}(x^*) H\ind(x^*) 
  dx^*
 }^2
  \right) ~ d\w
  \nn
  \\
&\le&
K
\int_\mathcal{B} 
 \sum_{m,\ell \in \mathcal C}  
 W^{-2}
\left( 
 \int_{r_+}^{R_1} 
  \abs{F}^2
  dr
+
\abs{
 \int_{R_1} ^{\infty}
  u_{out}(x^*) H\ind(x^*) 
  dx^*
 }^2
  \right) ~ d\w \nn.
\eea
This estimate may be integrated over $(r_0,r_1)$ to obtain
\be 
\label{eq:integralControl0}
\int_\mathcal{B} 
 \sum_{m,\ell \in \mathcal C}  
\int_{r_0}^{r_1}\abs{u\ind}^2 
~d\w
\le 
K
\int_\mathcal{B} 
 \sum_{m,\ell \in \mathcal C}  
 W^{-2}
\left( 
 \int_{r_+}^{R_1} 
  \abs{F}^2
  dr
+
\abs{
 \int_{R_1} ^{\infty}
  u_{out}(x^*) H\ind(x^*) 
  dx^*
 }^2
  \right) ~ d\w .
\ee 

The same argument, with \eqref{eq:representation} replaced with \eqref{eq:drepresentation} yields
\be 
\label{eq:integralControl1}
\int_\mathcal{B} 
 \sum_{m,\ell \in \mathcal C}  
\int_{r_0}^{r_1}\abs{(u\ind)' }^2 
~d\w
\le 
K
\int_\mathcal{B} 
 \sum_{m,\ell \in \mathcal C}  
 W^{-2}
\left( 
 \int_{r_+}^{R_1} 
  \abs{F}^2
  dr
+
\abs{
 \int_{R_1} ^{\infty}
  u_{out}(x^*) H\ind(x^*) 
  dx^*
 }^2
  \right) ~ d\w .
\ee

Collecting \eqref{eq:horizonControl}, \eqref{eq:infinityControl}, \eqref{eq:integralControl0} and \eqref{eq:integralControl1} and applying Theorem \ref{thm:quantModeStab} to control $W^{-2}$, we have
\bea
&& \int_\mathcal{B} 
 \sum_{m,\ell \in \mathcal C}  
 \left(
 \left(
\abs{u\ind(-\infty)}^2 + \abs{u\ind(\infty)}^2
 \right)
 +
\int_{r_0}^{r_1} \abs{\dd_{r^*}u\ind}^2 + \abs{u\ind}^2  ~ dr^*
\right)
~d\w
 \\
 &\le&
 K G
\int_\mathcal{B} 
 \sum_{m,\ell \in \mathcal C}  
 \left[
\abs{
 \int_{R_1} ^{\infty}
  u_{out}(x^*) H\ind(x^*) 
  dx^*
 }^2
 +
 \int_{r_+}^{R_1} 
  \abs{F}^2
  dr
+
  \w \int_{-\infty}^\infty Im(H\ind \bar u\ind) dr^* \right]
  ~ d\w.
\eea
It remains to control the right hand side of this estimate by  $\int_{\Sigma_0} \abs{\dd\psi}^2$.
The control of the first term is achieved using the proof of \cite[Lemma 3.3]{SR}. The remaining terms are controlled using the methods in \cite[\S 7]{C14-1}. 
\ep

\begin{remark}
We can replace the hyperboloidal hypersurface $\Sigma_0$ with an asymptotically flat hypersurface in Theorem \ref{thm:ILED} as follows. Let $\Sigma_0^*$ be an asymptotically flat hypersurface that agrees with $\Sigma_0$ for $\set{r \le R}$ and which lies in the past of $\Sigma_0$. Choosing $R$ large enough that $T$ is timelike in $\set{r \le R}$, applying the $T$ energy estimate immediately implies that
$$
\int_{\Sigma_0} \abs{\dd\psi}^2 \le C\int_{\Sigma_0^*} \abs{\nabla_{g_{\Sigma_0^*}} \psi }^2 + \abs{n_{\Sigma_0^*} \psi}^2,
$$
so we can then replace the right hand side of \eqref{eq:ILED} by 
this integral over an asymptotically flat hypersurface. 
\end{remark}

\bibliographystyle{plain}
\bibliography{allRefs}

\begin{thebibliography}{10}

\bibitem{AnBl}
Lars Andersson and Pieter Blue.
\newblock Hidden symmetries and decay for the wave equation on the {K}err
  spacetime.
\newblock {\em Pre-print}, http://arxiv.org/abs/0908.2265, 2009.

\bibitem{Ar11-2}
Stefanos Aretakis.
\newblock Decay of axisymmetric solutions of the wave equation on extreme
  {K}err backgrounds.
\newblock {\em J. Funct. Anal.}, 263(9):2770--2831, 2012.

\bibitem{Ar12}
Stefanos Aretakis.
\newblock Horizon instability of extremal black holes.
\newblock {\em ArXiv}, 1206.6598, 2012.

\bibitem{Ca68-2}
Brandon Carter.
\newblock Hamilton-{J}acobi and {S}chr\"odinger separable solutions of
  {E}instein's equations.
\newblock {\em Comm. Math. Phys.}, 10:280--310, 1968.

\bibitem{CK93}
Demetrios Christodoulou and Sergiu Klainerman.
\newblock {\em The global nonlinear stability of the {M}inkowski space},
  volume~41 of {\em Princeton Mathematical Series}.
\newblock Princeton University Press, Princeton, NJ, 1993.

\bibitem{C14-1}
Damon Civin.
\newblock {S}tability of subextremal {K}err--{N}ewman exterior spacetimes for
  linear scalar perturbations.
\newblock {\em To appear}, 2014.

\bibitem{DR09-1}
Mihalis Dafermos and Igor Rodnianski.
\newblock The red-shift effect and radiation decay on black hole spacetimes.
\newblock {\em Comm. Pure Appl. Math.}, 62(7):859--919, 2009.

\bibitem{DR10}
Mihalis Dafermos and Igor Rodnianski.
\newblock Decay for solutions of the wave equation on {K}err exterior
  space-times {I}-- {II}: The cases $\abs a \ll$ {M} or axisymmetry.
\newblock {\em arXiv}, 1010.5132, 2010.

\bibitem{DR12}
Mihalis Dafermos and Igor Rodnianski.
\newblock The black hole stability problem for linear scalar perturbations.
\newblock {\em Proceedings of the Twelfth Marcel Grossmann Meeting on General
  Relativity, T. Damour et al (ed.)}, 2011.

\bibitem{DR11}
Mihalis Dafermos and Igor Rodnianski.
\newblock A proof of the uniform boundedness of solutions to the wave equation
  on slowly rotating {K}err backgrounds.
\newblock {\em Invent. Math.}, 185(3):467--559, 2011.

\bibitem{DR08}
Mihalis Dafermos and Igor Rodnianski.
\newblock Lectures on black holes and linear waves.
\newblock In {\em Evolution equations}, volume~17 of {\em Clay Math. Proc.},
  pages 97--205. Amer. Math. Soc., Providence, RI, 2013.

\bibitem{DRS14}
Mihalis Dafermos, Igor Rodnianski, and Yakov Shlapentokh-Rothman.
\newblock Decay for solutions of the wave equation on {K}err exterior
  spacetimes {I}{I}{I}: The full subextremal case $\abs{a} < {M}$.
\newblock {\em Pre-print, http://arxiv.org/abs/1402.7034}, 2014.

\bibitem{Dy10}
Semyon Dyatlov.
\newblock Quasi-normal modes and exponential energy decay for the {K}err-de
  {S}itter black hole.
\newblock {\em Comm. Math. Phys.}, 306(1):119--163, 2011.

\bibitem{MR2014957}
Felix Finster, Niky Kamran, Joel Smoller, and Shing-Tung Yau.
\newblock The long-time dynamics of {D}irac particles in the {K}err-{N}ewman
  black hole geometry.
\newblock {\em Adv. Theor. Math. Phys.}, 7(1):25--52, 2003.

\bibitem{Ga12}
Oran Gannot.
\newblock {Q}uasinormal modes for {S}chwarzschild-{A}d{S} black holes:
  exponential convergence to the real axis.
\newblock {\em arXiv}, 1212:1907, 2012.

\bibitem{HS13-1}
Gustav Holzegel and Jacques Smulevici.
\newblock {Decay properties of {K}lein-{G}ordon fields on {K}err-{A}d{S}
  spacetimes.}
\newblock {\em Commun. Pure Appl. Math.}, 66(11):1751--1802, 2013.

\bibitem{HS13-2}
Gustav Holzegel and Jacques Smulevici.
\newblock Quasimodes and a lower bound on the uniform energy decay rate for
  {K}err-{A}d{S} spacetimes.
\newblock {\em arXiv}, 1303.5944, 2013.

\bibitem{K07}
Sergiu Klainerman.
\newblock Mathematical challenges of general relativity.
\newblock {\em Rend. Mat. Appl. (7)}, 27(2):105--122, 2007.

\bibitem{SR}
Yakov Shlapentokh-Rothman.
\newblock Quantitative mode stability on the real axis for the wave equation on
  the {K}err spacetime.
\newblock {\em To appear in Ann. Henri Poincar\'e}, 2013.

\bibitem{TaTo}
Daniel Tataru and Mihai Tohaneanu.
\newblock A local energy estimate on {K}err black hole backgrounds.
\newblock {\em Int. Math. Res. Not. IMRN}, 2:248--292, 2011.

\bibitem{Wa13}
Claude Warnick.
\newblock On quasinormal modes of asymptotically {A}nti-de {S}itter black
  holes.
\newblock {\em arXiv}, 1306.5760, 2013.

\bibitem{Wh89}
Bernard~F. Whiting.
\newblock Mode stability of the {K}err black hole.
\newblock {\em J. Math. Phys.}, 30(6):1301--1305, 1989.

\end{thebibliography}

\end{document}